\documentclass[prl,twocolumn]{revtex4}
\usepackage{graphicx}
\renewcommand{\v}[1]{{\bf #1}}

\newcommand{\s}{{\sigma}}

\def\eqa{\begin{eqnarray}}
\def\eea{\end{eqnarray}}
\newcommand{\eq}{\begin{equation}}
\newcommand{\ee}{\end{equation}}
\newcommand{\nn}{\nonumber\\}
\newcommand{\Eq}[1]{Eq.~(\ref{#1})}
\newcommand{\<}{\langle}
\renewcommand{\>}{\rangle}

\begin{document}

\title{Doped $t$-$J$ Model on a Triangular Lattice: Possible
Application to Na$_x$CoO$_2\cdot y$H$_2$0 and Na$_{1-x}$TiO$_2$}
\author{Qiang-Hua Wang$^{a}$,Dung-Hai Lee$^{b}$,
and Patrick A. Lee$^{c}$} \affiliation{${(a)}$ National Laboratory
of Solid State Microstructures,Institute for Solid State Physics,
Nanjing University, Nanjing 210093, China}
\affiliation{${(b)}$Department of Physics,University of California
at Berkeley, Berkeley, CA 94720, USA}
\affiliation{${(c)}$Department of Physics, Massachusetts Institute
of Technology, Cambridge, MA 02139, USA}


\begin{abstract}
We report the finding of time-reversal-symmetry-breaking
$d_{x^2-y^2}+ id_{xy}$ superconducting ground state in the
slave-boson mean-field theory for the t-J model on triangular
lattice. For $t/J=-5$ ($t/J=-9$) pairing exists for $x < 13\%$
($x<8\%$) upon electron doping, and $x <56\%$ ($x<13\%$) upon hole
doping. These results are potentially relevant to doped Mott
insulators Na$_x$CoO$_2$.yH$_2$O and Na$_{1-x}$TiO$_2$.

\end{abstract}

\maketitle

There is now broad agreement that the physics of the high T$_c$
cuprate is that of the doped Mott insulator. However, 17 years
after its discovery\cite{bednorz} the layered cuprates remain the
only materials which exhibit the phenomenon of high T$_c$
superconductivity.  There are three reasons which make the
cuprates unique: (1) The parent compound is a  Mott insulator with
$S={1\over 2}$ and no orbital degeneracy; (2) the structure is two
dimensional; and (3) the exchange energy $J$ is very large ($J
\approx 1500$ K).  Anderson\cite{anderson} has stressed the strong
quantum fluctuation of the $S={1\over 2}$ system in two
  dimensions. His resonating valence bond (RVB)
theory describes a liquid of spin singlet which  becomes a
superconductor when the holes are phase coherent. If this is a
general property of a  doped Mott insulator, it will clearly be
desirable to examine other examples which satisfy these three
criteria. Recently Takada {\it et al}.\cite{takada} reported the
discovery of  superconductivity in Na$_x$CoO$_2$.yH$_2$O with a
T$_c$ of 5 K  for $x$ = 0.35.  As these authors pointed out, this
system may be viewed as a Mott insulator with electron doping of
35\%.  The Co atoms are in a  triangular lattice and the Co$^{4+}$
atom is in a low spin $(S={1\over 2})$ state.  Thus this material
satisfies the first two criteria listed above. The value of $J$ is
not known at present, but the new discovery offers hope that a
second system which exhibits superconductivity by doping a  Mott
insulator may be realized. A summary of what is known about this
material and  a discussion in terms of RVB physics was given by
Baskaran.\cite{baskaran} In this paper we argue that the $t$-$J$
model on a triangular lattice is a reasonable starting point to
model these materials.  We estimate the value of $J$, and compute
the slave boson mean field phase diagram for both electron and
hole doping. We find $d_{x^2-y^2}+id_{xy}$ pairing over
significant range of doping for both $t/J=-5$ and $t/J=-9$. The
appearance of superconductivity shows an interesting particle-hole
asymmetry. We propose that Na$_{1-x}$TiO$_2$ may be an example of
the hole-doped system.

The transition metal oxide layers in
Na$_x$CoO$_2$ and Na$_{1-x}$TiO$_2$ form a common structure where
the transition metal is surrounded by an octahedral oxygen cage.
The cages are edge sharing, forming a layered structure.
The stacking of the oxide layers is different in the two materials and
we first describe the structure of Na$_{1-x}$TiO$_2$.
A way to
visualize the structure which is convenient for understanding the
electronic structure
\begin{figure}
\includegraphics[width=3.0in]{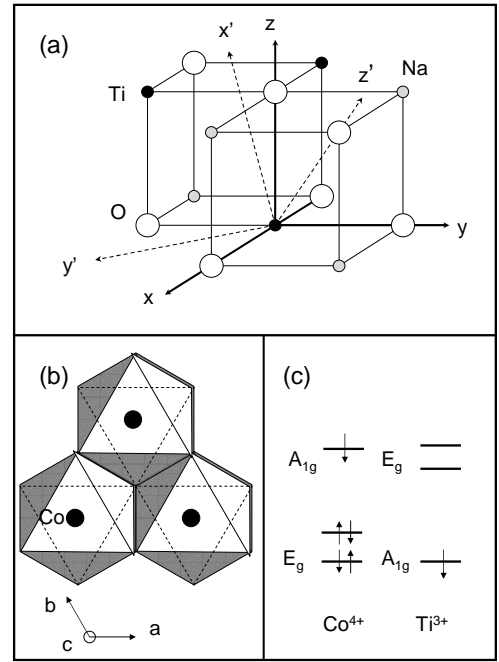}
\caption{(a) The NaTiO$_2$ structure.
(b) Projection of Fig.1(a) along the (1,1,1) direction showing the
oxide
layer.  This layer is similar in Na$_x$CoO$_2$ and NaTiO$_2$.
Oxygen occupies vertices of the solid (dashed) triangles which lie
above
(below) the cobalt or titanium layers.
(c) The splitting of the $t_{2g}$ levels due to the
distortion of the oxygen octahedra.}
\end{figure} is given in Fig.1(a) (adopted from Ref.\cite{sawatzky}).
It shows a piece of the crystal structure in a co-ordinate system
which is standard for the oxygen octahedron. The hexagonal layered
structure is visualized by looking down on this structure along
the (1,1,1) direction. Thus the axis $z^\prime$ becomes the
$c$-axis in  the layered hexagonal structure notation.  One can
see stacking of O-Ti-O or
O-Co-O hexagonal layers, highlighted in Fig.1(b),
separated by a Na layer.
In Na$_{1-x}$TiO$_2$, the Na sits directly over the oxygen in the
layer below Ti (vertices of the dashed
triangle), whereas in Na$_x$CoO$_2$, the Na site is on top of the Co.
The next CoO$_2$ layer has Co on top of Na
but the stacking order of the oxygen layer is reversed (solid
triangle now lies below Co) thereby doubling the
unit cell along $c$.  This difference in stacking does not affect the
electronic structure of the CoO$_2$ layers
and will be ignored from now on.

From Fig.1(a) it is clear that the $d$ states are split into
$t_{2g}$ and $e_g$ orbitals by the octahedral environment.  A
shift of the oxygen along the (1,1,1) direction splits the
$t_{2g}$ orbitals into\cite{sawatzky}  \eq A_{1g} = (d_{xy} +
d_{yz} + d_{zx})/ \sqrt{3} \ee and a doublet (labelled $E_g$ to
distinguish from $e_g$) \eq E_g = \Big((d_{zx} - d_{yz}) /
\sqrt{2}, (-2d_{xy} + d_{yz} + d_{zx})/ \sqrt{6}\Big).\ee  The
band calculation of Singh\cite{singh} showed that the splitting is
fairly large in Na$_{0.5}$CoO$_2$ with $A_{1g}$ lying higher than
$E_g$. As shown in  Fig.1(c), in Co$^{4+}$ the unpaired spin
occupies the non-degenerate $A_{1g}$ orbital.  We note that the
$t_{2g}$ orbitals have lobes which point to the mid-point of lines
connecting the oxygens in the octahedral cage. From Fig.1(a) and
Eq.(1) we see that the $A_{1g}$ orbitals on nearest neighbor Co
have components which point directly at each other. Thus unlike
the cuprate, where the hopping is via the Cu-O covalent band, in
the  cobalt compounds, the direct overlap between the $A_{1g}$
orbitals form a band. Due to band overlap, it is difficult to
extract the $A_{1g}$ bandwidth from Singh's calculations
\cite{singh} but we estimate it to be between 1 to 1.4 eV.  From
this we extract the hopping integral $t$ for the $t$-$J$ model \eq
H = -t\sum_{\< ij\>}\Big( P c_{i\sigma}^\dagger c_{j\sigma} P+{\rm
h.c.}\Big) + J \sum_{\< ij\>} \left( {\bf S}_i \cdot {\bf S}_j -
{1\over 4} n_in_j \right)\label{tj} \ee to be $t = -0.11$ to $-
0.15$ eV.  In \Eq{tj} the projection operator $P$ removes
double/zero occupancy for hole/electron doping respectively. We
note that the overlap between $A_{1g}$ orbitals is positive and
the negative sign of $t$ is a consequence of the sign convention
chosen in Eq.(3).  As emphasized by Baskaran,\cite{baskaran} there
is no particle-hole symmetry in the triangular lattice and the
sign of $t$ is important. Singh's band structure \cite{singh}
shows a maximum in the band structure at $\Gamma$, confirming the
negative sign of $t$.  The Fermi surface of Na$_{0.5}$CoO$_2$
consists of a hole pocket of area ${1\over 4}$ around the $\Gamma$
point.\cite{singh} This is consistent with photoemission
results.\cite{valla}

It is much more difficult to estimate the exchange constant $J =
4t^2/U$ because the $U$ parameter is highly uncertain due to
screening. We appeal to another $t_{2g}$ $S = {1\over 2}$ system
where $J$ has been determined experimentally. In TiOCl the
$t_{2g}$ orbital is orbitally ordered and forms one-dimensional $S
= {1\over 2}$ spin chains.\cite{seidel} The exchange interaction
was found to be 660 K by fitting the spin susceptibility  to the
Bonner-Fisher curve. In this case the $t_{2g}$ orbitals are pure
$d_{yz}$ and point directly at each other. The bandwidth from band
structure calculations is also about 1 eV, leading to $t \approx
-1/4$ eV. Note that $t$ for Na$_x$CoO$_2$ is considerably smaller
even though the Co-Co distance at 2.84 angstrom is shorter than
the Ti-Ti distance (3.38 angstrom). This is because only one
component out of three in Eq.(1) points directly towards each
other. Assuming that $U$ is similar, our best guess for  $J$ is 12
to 24 meV and $|t/J| \approx$ 6 to 9. Note that our estimate of
$J$ for Na$_x$CoO$_2$ is about an order of magnitude smaller than
that for the cuprates.

It is interesting to consider another $S = {1\over 2}$ system
which corresponds to single occupation of the $t_{2g}$ orbitals.
NaTiO$_2$ has similar layer structure as Na$_x$CoO$_2$ but the
Na layer is nominally fully occupied.  The Ti$^{3+}$ has $d^1$
configuration giving rise to unpaired $S = {1\over 2}$. Contrary
to Na$_{0.5}$CoO$_2$, the $A_{1g},E_g$ splitting is much smaller
than the bandwidth, according to LDA calculations.\cite{sawatzky}
However, an LDA+U calculation shows that the $A_{1g}$ orbital is
occupied preferentially, so that the single electron again
occupies  the nondegenerate $A_{1g}$ band.\cite{sawatzky}
Reduction of the Na occupation (Na$_{1-x}$TiO$_2$) corresponds to
doping by $x$ holes. The undoped system is of great interest
because it is one of  the few known examples of the $S = {1\over
2}$ triangular antiferromagnet. However, the control of Na
stoichiometry presents a serious materials challenge and
relatively few studies have been carried out to date.\cite{clark}

As we mentioned earlier the occupation constraint of \Eq{tj} is
different for the electron and hole doping. In the case of
electron doping we shall perform a particle-hole transformation
$c_{i,\sigma} \rightarrow c_{i,-\sigma}^\dagger$ so that the
constraint always means no double occupancy and the sign of $t$ is
reversed. Consequently $t<0$ for hole-doped Na$_{1-x}$TiO$_2$ and
$t>0$ for electron-doped Na$_x$CoO$_2$.

\parindent 10pt The starting point of our calculation is the following
U(1) slave-Boson mean-field Hamiltonian\cite{bl} for the t-J model
at hand, $H_{MF}=H_{nm}+H_m$ with \eqa
H_{nm}&&=-\sum_{\<ij\>\s}\Big[\Big(t\sqrt{x_i x_j} +{3J\over
8}\chi_{ij}^*\Big)f^\dagger_{i\s}f_{j\s}+{\rm
h.c.}\Big]\nn&&-{3J\over 8}
\sum_{\<ij\>\s\s'}\Big[\Delta^*_{ij}f_{i\s}f_{j\s'}\epsilon_{\s\s'}+
{\rm h.c.}\Big]-\sum_{i\s}\mu_in_{i\s};\nn
H_{m}&&=J\sum_{\<ij\>}(\v S_i\cdot \v m_j+\v S_j\cdot \v m_i).
\label{mf}\eea Here $x_i=1-\sum_\s \<n_{i\s}\>$ and $\mu_i$ is the
corresponding local Lagrange multiplier. The doping level is given
by $x=\sum_i x_i/N$ with $N$ the lattice sites. The mean field
order parameters are $\chi_{ij}=\sum_{\s}\<f^+_{i\s}f_{j\s}\>$,
$\Delta_{ij}=\sum_{\s\s'}\epsilon_{\s\s'}\<f_{i\s}f_{j\s'}\>$, and
$\v m_i=\<\v S_i\>$ with $\v S_{i}=\sum_{ss'}\v
\s_{ss'}f^\dagger_{is}f_{is'}/2$. Note that a general
non-collinear magnetic order is allowed.

Our search for the mean-field solution proceeds in two stages.
First we assume the density to be uniform and perform an
unrestricted search of the parameters $\chi_{ij},\Delta_{ij}$ and
$\v m_i$ (over $6 \times 6$ elementary unit cells in a large
lattice). For $x < x_c$ we find $\chi_{ij}$ to be a real constant
as long as $x$ is nonzero and $\Delta_{ij}=0$ while the spins form
the $\sqrt{3} \times \sqrt{3}$ non-collinear structure identified
by Huse and Elser\cite{huse} for $x=0$.  If we denote the lattice
by $\v{r}_i = n\v{a} + l\v{b}$, $\v{a}  = \hat{x}$ and $\v{b} =
-{1\over 2} \hat{x} + {\sqrt{3}\over 2}\hat{y}$, then a
representative solution has $m^z_i=0$ and $m_i^x+im^y_i=|\v m|
e^{i\v Q\cdot\v r_i}$, where $\v Q=(2\pi/3,2\pi/\sqrt{3})$ and
$\v{Q}\cdot \v{r}_i = {2\pi \over 3}(n+l)$. $|\v m|={1 \over 2}$
at zero doping, and decreases gradually with increasing doping. It
is roughly $0.35 \sim 0.38$ (depending on our choices of $t/J$) at
$x=x_c$, where the long-range ordered magnetic state undergoes a
first order phase transition into a uniform $d_{x^2-y^2}\pm
id_{xy}$ superconducting state. The value of $x_c$ is
$3.8\%,3.7\%, 2.2\%,2.1\%$ for $t/J=5,-5,9,-9$ respectively. Even
if the pairing channel is removed, the magnetic order extends only
up to $x_*=4.5\%,\,4.2\%,\,2.4\%,\,2.4\%$ for the same set of
$t/J$. This is significantly different from the mean-field
solution of the square lattice with $t/J=3$ as appropriate for the
cuprates. In that case the system is inhomogeneous with
alternating stripes of superconducting and magnetic regions at
roughly $2\%\le x\le 12\%$, and the magnetic order becomes
commensurate with coexisting uniform pairing order at higher
doping levels up to $x_c\sim 20\%$. \cite{hanwanglee} For the
hexagonal lattice, the stable region of the anti-ferromagnetic
solution is greatly reduced. This is reasonable in view of the
frustration of the lattice, which disfavors magnetic order and was
the original motivation of the RVB as a competing
state\cite{anderson}.

\begin{figure}
\includegraphics[width=8.5cm]{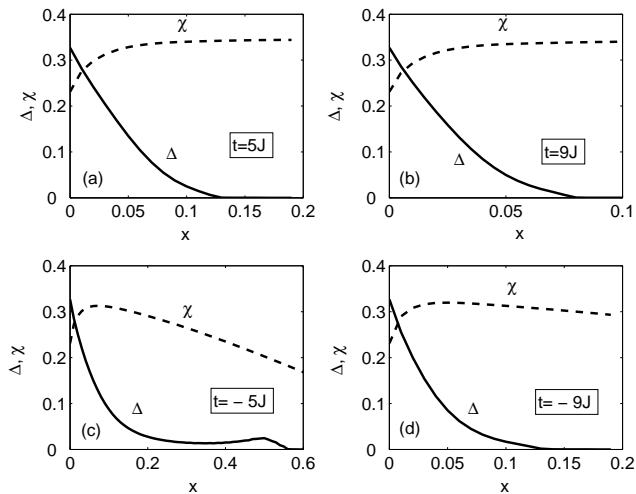}
\caption{Dependence of the order parameters on hole doping ($x$):
$t=5J$ (a), $9J$ (b), $-5J$ (c), and $-9J$ (d). Note the scale
change in 2(b) and 2(c). }
\end{figure}

In the second stage of the calculation we concentrate on doping
greater than $x_c$. In view of the first-order nature of the
transition to magnetic order, we shall present results without
magnetic order even for $x < x_c$, with the understanding that the
solution is only locally but not globally stable. We suppress
magnetic order and perform an unrestricted search for solutions of
$H_{nm}$ in Eq. (4) on a $400\times 400$ lattice. In the
elementary unit cell (varied from $1\times 1$ up to $4\times 8$)
we put random initial values for $\chi_{ij}$, $\Delta_{ij}$, $x_i$
and $\mu_i$, and evolve them so that the mean-field free energy is
minimized at the desired doping level. Note that this time we
allow nonuniform charge density. However, with our choice of
parameters, we found spatially non-uniform solutions only for
$x<x_u$, where $x_u$ is always less than $x_c$ and these will be
ignored from now on. Due to the uncertainty of $t/J$ we have
studied two cases $|t/J|=5$ and $|t/J|=9$. It turns out that in
both cases there is a significant doping range in which the ground
state is superconducting. The pairing symmetry is always
$d_{x^2-y^2}+ id_{xy}$, hence breaks the time reversal symmetry.
It is interesting to note that if one ignores the magnetic order
parameter at zero doping the same mean-field theory predicts a
degenerate family of ground states. Among them the $d_{x^2-y^2}+
id_{xy}$ paired state and the $\pi/2$ flux state are two
examples.\cite{laughlin,tk} To visualize the pairing pattern,
consider an arbitrary site. The $\Delta_{ij}$ associated with the
six bonds stemming from it has the form $\Delta_{ij} =\Delta
e^{i2\theta_{ij}}$ for $d_{x^2-y^2}+id_{xy}$ pairing and
$\Delta_{ij} =\Delta e^{-i2\theta_{ij}}$ for $d_{x^2-y^2}-id_{xy}$
pairing. To generate the pairing field over the entire lattice,
simply translate the above pattern to other sites. In the absence
of a magnetic field the $d_{x^2-y^2}+id_{xy}$ and
$d_{x^2-y^2}-id_{xy}$ pairing are degenerate.


The doping dependence of the zero temperature pairing order
parameter $\Delta=|\Delta_{ij}|$ is plotted in Figs.2 (solid
lines) for $t=5J$ (a), $9J$ (b), $-5J$ (c), and $-9J$ (d), where
we observe that pairing exists for $x\le 13\% $ (a), $8\%$ (b),
$56\%$ (c) and $13\%$ (d). The peak in $\Delta$ at $x=0.5$ in
Fig.2(c) is explained by the Van Hove peak in the free electron
density of states. Also shown in Figs.2 is the hopping order
parameter $\chi=|\chi_{ij}|$ (dashed lines). It is weakly
dependent on $x$ for $x<30\%$. Interestingly $\chi_{ij}$ has the
same sign of $t$ from our calculation. This is a reasonable result
as it increases the band width of the fermions so that the kinetic
energy is lowered. Finally, a much more important feature in
Figs.2 is the considerable asymmetry between electron doping
($t>0$) and hole doping ($t<0$). This is due to the particle-hole
asymmetry in the free-electron dispersion on the triangular
lattice. Indeed, the stronger pairing for hole doping is due to
the increases in the Fermi level density of state as the averaged
occupation decreases from half-filling.

We have also computed the onset temperature of the RVB fermion
pairing and Bose-Einstein condensation (BEC) of holons.\cite{note}
The mean-field superconducting transition is the smaller of the
RVB and BEC curves (Figs.3). As usual, the Bose condensation
temperature is an overestimate of the phase coherence temperature
of the slave bosons. More generally because of the proximity to
the Mott insulator limit, we expect that the superconducting
transition temperature will be determined by the superfluid
density at low doping, and by the onset of pairing at higher
doping, forming a phase diagram similar to that of the cuprates.
It is worth noting that in the case of cuprates, the thermal
excitation of nodal quasi-particles significantly reduces the
superfluid density and therefore the transition temperature at low
doping.\cite{pal,corson} The existence of a full gap in the
$d_{x^2-y^2}+id_{xy}$ state suppresses this possibility and
everything else being equal, we can expect a higher $T_c$ in this
case.

\begin{figure}
\includegraphics[width=8.5cm]{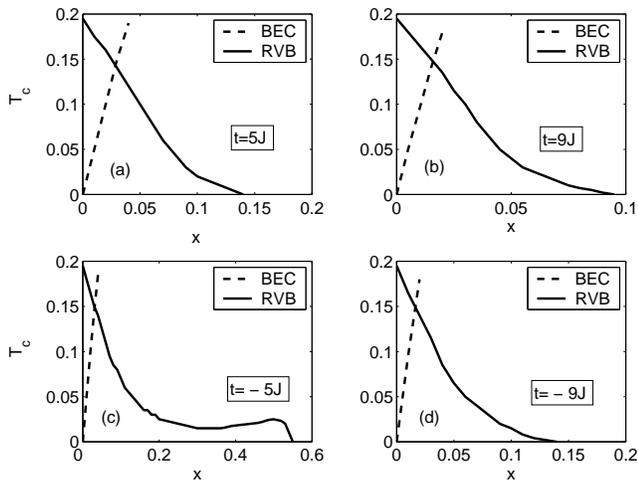}
\caption{Dependence of the RVB and BEC critical temperatures (in
units of J) on hole doping ($x$): $t=5J$ (a), $9J$ (b), $-5J$ (c),
and $-9J$ (d). Note the scale change in 2(b) and 2(c). }
\end{figure}

We conclude that within the slave boson mean field theory, the
systems exhibits non-colinear antiferromagnetism for $x<x_c$.
Above this lower critical concentration and below certain upper
critical concentration the system exhibit time reversal symmetry
breaking $d_{x^2-y^2}+ id_{xy}$ pairing state. The orbital moment
produces a magnetic field which can be detected by $\mu$SR. The
field has been estimated to be 15 Gauss for the anyon model
\cite{halperin} but a calculation based on the slave boson mean
field theory yields a smaller estimate of 1 Gauss near a vacancy.
\cite{blw}
If the orbital moment in neighboring layers are parallel,
this state has been predicted to exhibit fascinating new effects
such as quantized spin Hall conductance and anomalous Hall thermal
conductivity.\cite{hall} The orbital moment corresponds to roughly
${1\over 20}\mu_B$, and field cooling in a modest magnetic field
may line up the orbital moments. It will clearly be desirable to
achieve such a state in the laboratory. For $|t/J|=5$ ($|t/J|=9$),
the $d_{x^2-y^2} + id_{xy}$ superconductor is expected to exist
over a low doping range $x < 13\%$ ($x<8\%$) for electron doping,
and a wider range $x <56\%$ ($x<13\%$) for hole doping. Thus some
ingredient in addition to the t-J model may be needed to explain
the experimental observation of superconductivity at $x=0.35$ in
electron-doped Na$_x$CoO$_2\cdot y$H$_2$O. Exploration of the
Na$_{1-x}$TiO$_2$ system may be more promising. We also note that
Na$_{0.7}$CoO$_2$ is a metal which exhibits unusual behavior such
as linear $T$ resistivity and large magnetic-field-dependent
thermal power.\cite{tp} This is also inconsistent with the mean
field prediction of Fermi liquid in the overdoped region,
suggesting that some additional physics may be at work. In any
event, the new observation opens up the possibility of changing
the doping concentration in a controlled way by intercalation and
much new physics surely remains to be discovered.

After the completion of this work we have seen a paper by Kumar
and Shastry \cite{kumar} reaching similar conclusions. However, we
do not agree with their identification of the sign of $t$.

\acknowledgments{We thank Fangcheng Chou, Joel Moore, T.K. Ng, and
N.P. Ong for helpful discussions. QHW is supported by NSFC
10204011 and 10021001, and by the Ministry of Science and
Technology of China (NKBRSF-G1999064602). He also thanks Z. D.
Wang for hospitality in the University of Hong Kong. DHL is
supported by DOE grant DE-AC03-76SF00098. PAL is supported by NSF
DMR-0201069. He also thanks the Miller Institute at Berkeley for
support.}

\end{document}